\documentclass[aps,prl,showpacs,twocolumn,amsmath,amssymb]{revtex4-1}

\usepackage{graphicx,graphics,color,epsfig}
\usepackage{bm}
\usepackage{amsmath,amssymb,amsfonts}
\usepackage{epstopdf}
\usepackage{dcolumn}
\usepackage{times,xspace}

\begin{document}


\title{Role of Fermi surface anisotropy in the study of gap anisotropy using magnetic-field-angle dependence of thermal oscillations in $A_y$Fe$_{2-x}$Se$_2$ superconductors}

\author{Tanmoy Das$^1$}

\author{Anton B.~Vorontsov$^2$}

\author{Ilya Vekhter$^3$}

\author{Matthias J.~Graf$^1$}
\affiliation{$^1$ Theoretical Division, Los Alamos National Laboratory, Los Alamos, New Mexico 87545, USA\\
$^2$ Department of Physics, Montana State University, Bozeman, Montana 59717, USA\\
$^3$ Department of Physics and Astronomy, Louisiana State University, Baton Rouge, Louisiana 70803, USA}

\date{\today}

\begin{abstract}
We present a numerical study of the field-angle resolved oscillations of the thermal conductivity and specific heat under rotated magnetic field in the $A_y$Fe$_{2-x}$Se$_2$ [$A$=K,Rb,Cs,(Tl,K)] superconductors, using realistic two-band Fermi surface parameterization. Our key finding is that even for isotropic pairing on an anisotropic Fermi surface, the thermodynamic quantities exhibit substantial oscillatory behavior in the superconducting state, even much below the upper critical field. Furthermore, in multiband systems the competition of anisotropies between two Fermi surfaces can cause a double sign reversal of  oscillations as a function of temperature, irrespective of gap anisotropy. Our findings put severe constraints on simple interpretations of field-angle resolved measurements widely used to identify the angular structure of the superconducting gap.
\end{abstract}

\maketitle


The identification of the symmetry of the superconducting (SC) order parameter is an important step toward unraveling the pairing mechanism in any novel superconductor. For iron pnictides, the presence of hole and electron pockets at the $\Gamma$ and M points has led to the proposal of $s^{\pm}$-wave pairing \cite{mazinpnictide,kuroki,chubukov,bang} due to interband nesting between them. However, the recent discovery of the layered high-temperature superconductors $A_y$Fe$_{2-x}$Se$_2$, with $A$=K,Rb,Cs,(Tl,K), has challenged the consensus for the pairing symmetry and mechanism of superconductivity in this class of materials \cite{FeSe}. The iron-selenide family has a crystal structure similar to the iron-pnictide material BaFe$_2$As$_2$, but with hole pockets eliminated completely from the Fermi surface (FS) at the $\Gamma$ point in the Brillouin zone, yet the SC transition temperature $T_c$ is  comparable to that of iron pnictides. Various theoretical proposals have been put forward which support either the survival of $s$-wave pairing \cite{mazin,JianXin,Fang2011}, the emergence of nodal $d$-wave gap \cite{nodal_dwave,Maiti2011}, or more popularly nodeless $d$-wave gap \cite{maier,DHLee,Das_FeSe,Dasmodulated}. Indirect experimental evidence suggests isotropic pairing symmetry \cite{FeSenodeless1,FeSenodeless2,FeSenodeless3}, consistent with isotropic gaps reported in angle-resolved photoemission spectra \cite{HDing,Xu2012}. Therefore, direct high-precision imaging of the structure of the gap function and the location of the nodes, if they exist, is required. An effective and accurate technique for measuring the angular structure of the bulk gap relies on probing thermodynamic properties in a rotating in-plane magnetic field. For cuprate, pnictide, and heavy-fermion superconductors, this technique has been used widely to identify the SC pairing symmetry by mapping the field-angle dependence of the thermal conductivity or specific heat onto the angular structure of the SC gap and its pairing symmetry \cite{Matsuda,Miranovic,Yu1995,Aubin1997,Izawa2001,Park2003,Aoki2004,Deguchi2004,Park2008,An2010,Machida2012,Anton}.

In this Letter, we demonstrate that detailed knowledge of the FS topology and parameters is necessary for relating the nature of the oscillations to the nodes or minima of the gap structure. This is especially important for materials where the FS anisotropy is substantial, as is the case in layered iron selenides. To be quantitative and unambiguous about the shape of the SC gap, it is required to incorporate realistic FS topology, Fermi  velocities, and density of states (DOS) at the Fermi level into self-consistent calculations of thermal properties.

To accomplish this goal, we focus on layered iron-selenide superconductors and study the information embedded in the angle-resolved specific heat coefficient, $\gamma=C/T$, and thermal conductivity,  $\kappa$, in a rotating in-plane magnetic field using realistic tight-binding dispersions derived from first-principles electronic structure calculations. The main results of our calculations are: (1) For purely isotropic pairing symmetry, moderate FS anisotropies of layered iron-selenide superconductors are sufficient to introduce field-angle-dependent oscillations in the specific heat and thermal conductivity over a significant range of temperatures and at intermediate to high magnetic fields in the SC state. We find an inversion of the oscillation pattern as a function of temperature, which shows that oscillations are not a simple consequence of the anisotropy of the upper critical field. Therefore not all such oscillations at intermediate fields can be taken as proof of strong anisotropy in the SC gap. (2) For isotropic gaps on the FSs, the oscillations in $\gamma$ may change sign once or twice as a function of temperature. We identify the out-of-phase FS anisotropies between bands as the source for two sign reversals. (3) Complex field-angle dependence of the specific heat and thermal conductivity  for anisotropic FSs suggests that comparison of both quantities with material-specific theories is needed to identify the pairing symmetry and gap structure.

\begin{figure}[top]
\rotatebox[origin=c]{0}{\includegraphics[width=0.90\columnwidth]{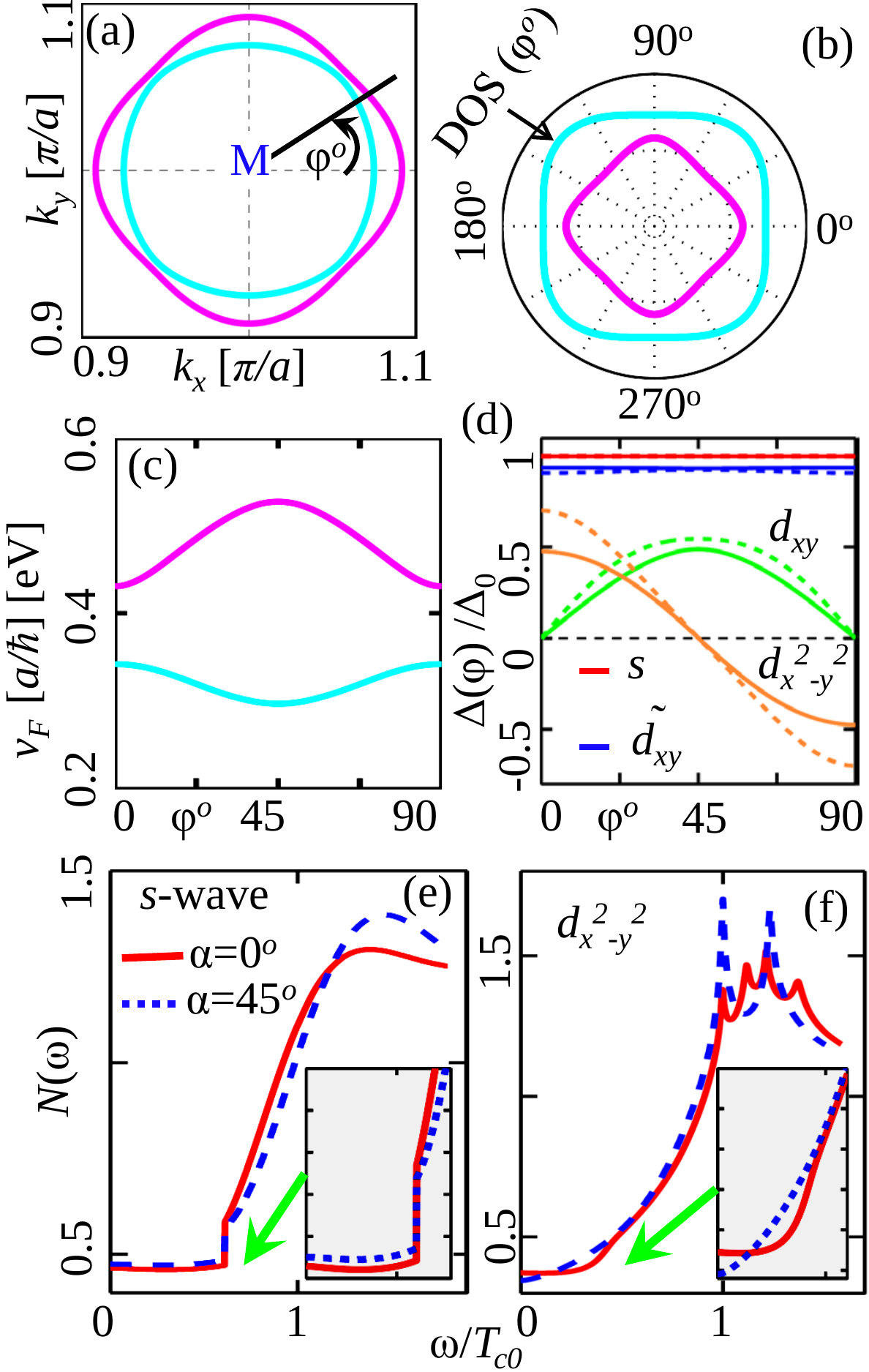}}
\caption{(Color online) (a) Different colors show different electron-pockets at $k_z=0$ within the tight-binding model of the 2-Fe unit cell \cite{Das_FeSe}.
(b) Polar plots of computed normal-state DOS at the Fermi level show the out-of-phase anisotropy between bands (same color as in (a)). (c) The tight-binding in-plane Fermi velocities at $k_z=0$  vs.\  azimuthal FS angle $\phi$ exhibit out-of-phase FS anisotropy in units of lattice parameter $a$.
(d) Gap functions  $\Delta_n(\phi)$ on FS1 (solid) and FS2 (dashed) for pairing symmetries considered. The nodeless states have negligible FS anisotropy. (e)-(f) Field-induced total SC DOS at $T=0$ vs.\  energy at two representative field angles $\alpha$ for $s$ and $d_{x^2-y^2}$ pairing. Note the low-energy crossings in the DOS (arrows) related to the low-$T$ sign reversals in the oscillations of $\gamma$ and $\kappa$; we used $H/H_{c2}=0.5$ for $s$ wave and 0.1 for  $d_{x^2-y^2}$ wave.}
\end{figure}

{\it Anisotropy in FS and SC gap.$-$}
In the iron selenides the Fe vacancy completely eliminates the hole pocket at the $\Gamma$ point, and the FS consists of two concentric electron pockets at the $M$ point in the 2-Fe unit cell. This picture follows from first-principles calculations \cite{LDA}  and photoemission spectroscopy \cite{HDing,Xu2012}. Here, we use a first-principles derived tight-binding parameterization of the electronic dispersion \cite{Das_FeSe} with a weak $k_z$ dispersion as input to obtain all necessary FS parameters for a self-consistent transport calculation. Cuts of the corresponding FSs are shown in Fig.~1(a) with calculated normal-state DOS in 1(b) and moderately anisotropic Fermi velocities in 1(c). These figures demonstrate the out-of-phase in-plane anisotropies of the FS parameters on the electron pockets at the M points.

We consider three nodeless gaps with $s$, $s^{\pm}=\cos{k_xa}+\cos{k_ya}$, and extended ${\tilde d}_{xy}=\sin{(k_xa/2)}\sin{(k_ya/2)}$ symmetry, shown in Fig.~1(d). Since the FSs are centered around M=($\pi,\pi,0$), and its equivalents, all three including ${\tilde d}_{xy}$ are nodeless on the FS~\cite{Das_FeSe}. As all nodeless gaps exhibit very similar behavior, and thus we show detailed results only for the isotropic pairings $s$. For the nodal SC gaps, we consider two pairings symmetries as $d_{x^2-y^2}=\cos{k_xa}-\cos{k_ya}$ and $d_{xy}=\sin{k_xa}\sin{k_ya}$, with detailed results presented for $d_{x^2-y^2}$ pairing. The gap structure for each pairing on the FSs is demonstrated in details in the supplementary material (SM) \cite{SM}.

{\it Brandt-Pesch-Tewordt (BPT) approximation.$-$}
We solve the quasiclassical Eilenberger equation within the extended BPT approximation
\cite{BPT,WPesch1975,AHoughton1998,Vekhter99,Anton,Anton2010,AntonI,AntonII} to solve for the field-angle induced SC DOS, $N_n(\omega, {\bm k_f}; {\bm H})$,  together with the self-consistency equations for the SC order parameter, $\Delta_n({\bm k_f};{\bm  H})$, and transport lifetime, $\tau_{n}(\omega,{\bm k_f};{\bm  H})$. Here ${\bm H}$ is the magnetic field applied at angle $\alpha$ with respect to the (100) direction, $\bm k_f$ is Fermi momenta and
$n=1,2$ is the band index. The transport lifetimes encodes the combined effects of impurity and vortex scattering. The BPT approximation implies that the DOS is obtained by averaging the normal quasiparticle Green's function over the unit cell of the Abrikosov vortex lattice. This produces quantitatively correct results near the upper critical field over the range $0.5 H_{c2} \lesssim H < H_{c2}$ \cite{BPT,Brandt,Delireu} for isotropic gap, but it extends to low fields for nodal and strongly anisotropic gaps \cite{AHoughton1998,TDahm2002,Anton}. The SC gaps are evaluated by solving the coupled BCS gap equations for $\Delta_n({\bm k_f}; {\bm H})$ at each applied field ${\bm H}$. We simplify the problem by considering interband pairing only and eliminate the pairing potential in favor of the bare transition temperature $T_{c0}$ within weak-coupling theory~\cite{Anton2010}. Details of the calculations are given in the SM \cite{SM}.

Based on the above-mentioned self-consistent solutions, the specific heat, $C$, and thermal conductivity, $\kappa$, are computed numerically from the solution of the Eilenberger equations~\cite{AntonI,AntonII}. However, to get a qualitative understanding of how the interplay of FS anisotropy and gap anisotropy contribute to the results of $C=C_1+C_2$ and $\kappa=\kappa_1+\kappa_2$, we write down the approximate low-$T$ expressions:
\begin{eqnarray}
&&C_n(\alpha) \! \approx \!\!
\int_{-\infty}^\infty \!
d\omega \frac{\omega^2 \langle N_n(\omega, {\bm k_f}; {\bm H}) \rangle_{FS}}{4 T^2 {\rm cosh}(\omega/2T)^2} ,
\label{eq:C}
\\
&&\kappa_{n}^{xx}(\alpha) \! \approx \!\!
\int_{-\infty}^\infty \!\!
d\omega \frac{ \omega^2 \langle v_{n}^x({\bm k_f})^2 N_n(\omega, {\bm k_f}; {\bm H})\tau_{n}(\omega,{\bm k_f};{\bm  H}) \rangle_{FS} }{2 T^2 {\rm cosh}(\omega/2T)^2}.
\nonumber\\
\end{eqnarray}
The angle-dependent SC DOS is given by the unit-cell averaged quasiclassical retarded Green's function $g_n$: $N_n(\omega, {\bm k_f};  {\bm H})=-N_{fn}({\bm k_f}){\rm Im}~g_n(\omega,{\bm k_f}; {\bm H})/\pi$. 
Here $\langle \dots \rangle_{FS}$ stands for the FS integrals and ${\bm v}_{n}$ is the Fermi velocity in each band, see Fig.~1(c). In the normal state the DOS becomes $N_n(\omega, {\bm k_f}; {\bm H}) =N_{fn}({\bm k_f}) \sim 1/|{\bm v}_{n}({\bm k_f})|$ and $\tau_{n}(\omega, {\bm k_f}; {\bm H} )= \tau_{imp}$ \cite{impurity}.

The dominant contribution to the anisotropy in $C_n$ at low $T$ originates from the anisotropy in the SC DOS at $\omega= 0$,
\begin{equation}
  \langle N_n(0, {\bm k_f}; {\bm H}) \rangle_{FS}\approx
  \left\langle
  \frac{ N_{fn}({\bm k_f})  }{
   \sqrt{1+
  \left(\frac{2\Lambda \widetilde\Delta_n({\bm k_f}; {\bm H})}{{|{\bm v}_{n}^\perp({\bm k_f}; {\bm H})}|}\right)^2}
  }
  \right\rangle_{FS} ,
  \label{DOS}
\end{equation}
where $\Lambda=\left(\hbar c/2|e| H\right)^{1/2}$ is the magnetic length of order the coherence length $\xi$ between $0.5 H_{c2} < H < H_{c2}$, $|{\bm v}_{n}^\perp|$ is the component of the rescaled Fermi velocity normal to ${\bm H}$, and $\widetilde\Delta_n$ is the impurity renormalized order parameter~\cite{AntonI}. For a cylindrically symmetric FS, the angle-dependence of $|{\bm v}_{n}^\perp|$ is determined solely by the field direction~\cite{AntonI}, its interplay with the profile of $\widetilde\Delta_n({\bm k_f}; {\bm H})$ gives the anisotropy of $C_n(\alpha)$. For complex FSs, there is an additional weighting of the integral due to momentum-dependence of $N_{fn}({\bm k_f})$ and ${\bm v}_{n}({\bm k_f})$, leading to strong field-angle oscillations even for an isotropic gap.

Calculations for single-band models~\cite{Anton,AntonI,UdagawaI,MiranovicI} and experiments on several classes of materials~\cite{An2010,FeSeTe_exp,cerhin5}  demonstrated that the anisotropy in heat capacity
undergoes inversion as $T$ and $H$ change. We qualititively reproduce the general sign reversal of the oscillations for nodal pairings even after replacing quasi-cylindrical FSs with more realistic and material-specific FS anisotropies.
In addition, for an {\it isotropic} gap, we find that even for a single strongly anisotropic FS, one or more inversion(s) of the oscillation can occur.  Multiband effects add additional complexity due to competing FS anisotropies and self-consistently evaluated multiple gap amplitudes, and the intuitive one-to-one mapping between oscillations and nodal directions becomes easily lost at finite temperatures and fields.

\begin{figure}[top]
\rotatebox[origin=c]{0}{\includegraphics[width=0.99\columnwidth]{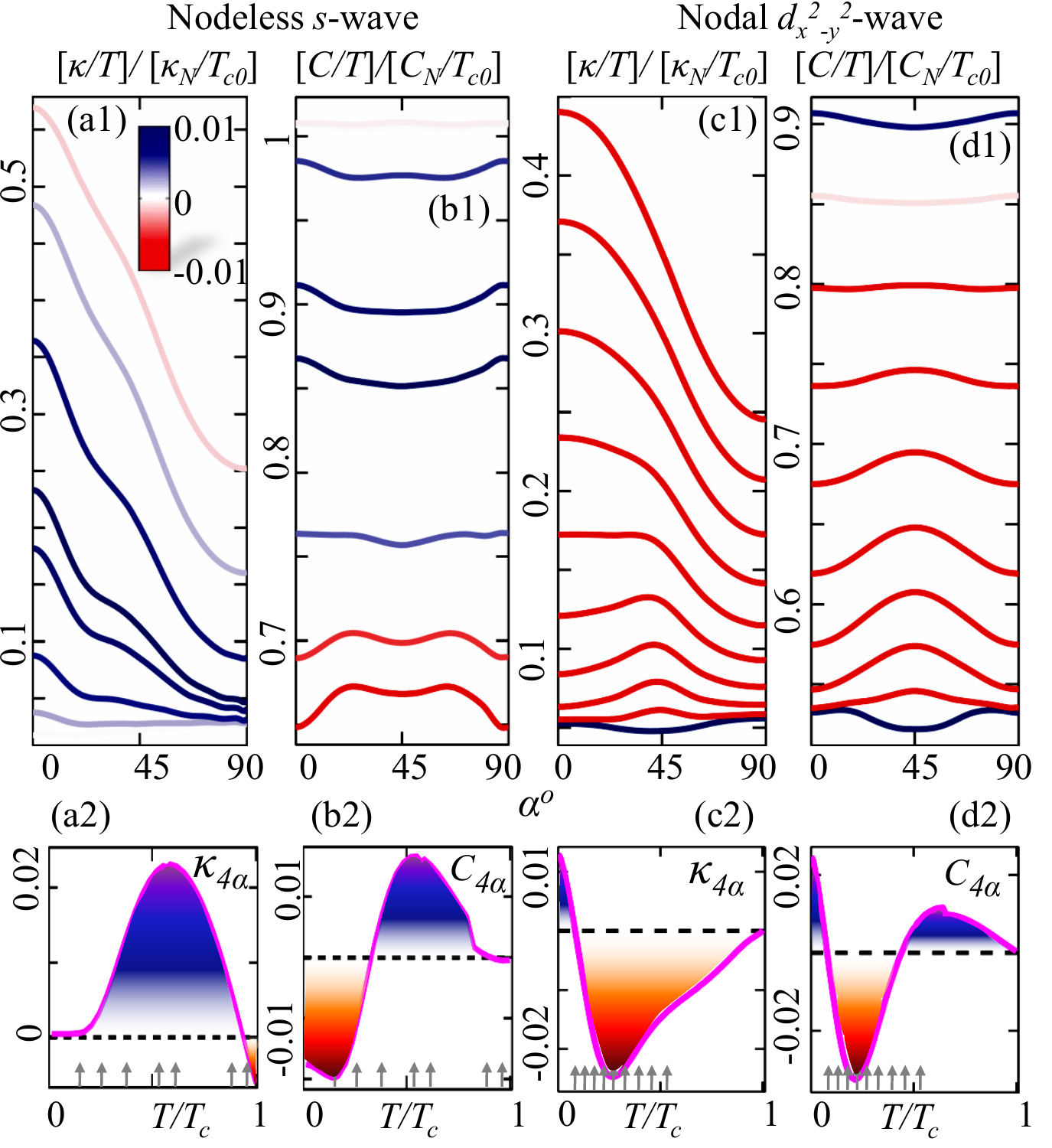}}
\caption{(Color online) (a1) Normalized thermal conductivity  $\kappa/T$ along (100) direction (normalized to its normal-state value $\kappa_N/T_c$) as a function of in-plane field-angle $\alpha$ at fixed $H/H_{c2}=0.5$ for $s$ wave,
plotted from low to high $T$ (bottom to top curves). Each subsequent curve is shifted vertically by 0.01 for clarity.
Each curve is colored by the amplitude of the fourfold oscillation given in panel (a2);
a uniform color map is used for values below $-0.01$ and above $0.01$.
(a2) The fourfold  amplitude of $\kappa/T$ is plotted as a function of $T$.
The vertical arrows in the bottom row depict the temperatures at which curves in top panels are shown.
Panels (b1)-(b2): Normalized specific heat for same parameters as in (a1)-(a2).
Panels (a) and (b) are for purely isotropic $s$-wave case,
while similar plots in (c) and (d) are shown for nodal $d_{x^2-y^2}$-wave symmetry at $H/H_{c2}=0.1$.}
 \label{fig2}
\end{figure}

{\it Results and discussions.$-$} Figs.~1(e)-1(f) show the field-induced SC DOS as a function of quasiparticle energy below the SC gap for $\alpha=0^o$ and $\alpha=45^o$ for $s$ and $d_{x^2-y^2}$ OPs. We immediately see that the difference between SC DOS at these two representative angles changes sign at finite $\omega$ for both cases, opening the possibility for the low-$T$ sign reversal of the oscillations in the specific heat as a function of temperature.

We present the full angle-dependent profiles of $C(\alpha)$ and $\kappa (\alpha)$ for several temperatures at a representative low field for an isotropic $s$-wave gap (at $H/H_{c2}=0.5$) and a nodal $d_{x^2-y^2}$ gap (at $H/H_{c2}=0.1$) in Fig.~2. It is interesting to note that there is a prominent angle dependence for the $s$-wave gap (Figs.~2(a) and 2(b)), whose nature is far more complex than what can be interpreted by conventional harmonics of pairing symmetries. Especially, at low $T$ the peak position of $C(\alpha)$  is shifted from high-symmetry values and lies somewhere between $\alpha=0^o$ to $45^o$, see Fig.~2(b1). Such complex field-angle dependence is a manifestation of the out-of-phase anisotropies on both FSs, shown in Fig.~1(a)-1(c). For nodal $d_{x^2-y^2}-$pairing, the behavior of  oscillations of $C(\alpha)$ and $\kappa(\alpha)$ is similar to  results obtained for quasi-cylindrical FSs \cite{Anton2010}, however the amplitude of  oscillations and the location of  sign reversals  are modified \cite{cecoin5}.

We estimate the amplitudes of the fourfold oscillations by defining $C_{4\alpha} (T)=\Pi_0^C-\Pi_{45}^C$, where $\Pi_{\alpha}^C=[C(\alpha,T)/T]/[C_N/T_c]$ and
$\kappa_{4\alpha}(T)=[\Pi_0^\kappa+\Pi_{90}^\kappa]/2-\Pi_{45}^\kappa$, where $\Pi_{\alpha}^\kappa=[\kappa^{xx}(\alpha,T)/T]/[\kappa^{xx}_N/T_c]$,
and $C_N$ and $\kappa_N$ are their corresponding normal-state values at $T_c$ \cite{kappa}.
Such definition removes any twofold contribution from $\kappa$.
The corresponding results are plotted in the lower panels of Fig.~2. We obtain several sign reversals in $C_{4\alpha} (T)$ and $\kappa_{4\alpha}(T)$ for both isotropic $s$-wave and nodal $d$-wave gaps. Earlier such sign-reversal feature was only found for highly anisotropic or nodal gap structure \cite{Anton,AntonI,Anton2010,An2010,FeSeTe_exp,cerhin5}. For this realistic FS parameterization of the double layered iron selenide, we find indications of two sign reversals even for the $s$-wave gap. Although the second sign reversal at high $T$ may be difficult to discern, it is visible in Figs.~2(b1) and 2(b2) and as a white region in Fig.~3(a1). We verified that the magnitude of oscillations depends on the out-of-plane electronic hopping, defined in SM\cite{SM}.
In addition, our calculations show that the amplitude of fourfold oscillations for $s$-wave pairing is roughly half of that for nodal pairing.

It is noteworthy that the $T$-dependence of the fourfold oscillations in Figs.~2(b2) and 2(d2) reflects on the energy dependence of the SC DOS in Figs.~1(e)-1(f) at the same value of  $H$. For example, for $s$-wave gap $C(\alpha=45^o)> C(\alpha=0^o)$ at low $T$ in Fig.~2(b2), which corresponds to $N(\alpha=45^o) > N(\alpha=0^o)$ at low energy in Fig.~1(e). The opposite anisotropy in $C(\alpha)$ at high $T$ corresponds to the inversion of the DOS anisotropy at higher energies. We conclude that the anisotropy in $C(\alpha)$ for isotropic $s$-wave pairing is not merely a manifestation of the anisotropy in $H_{c2}$ (irrelevant at low fields, since it is tied to the FS shape), but is a reflection of
the field-induced spectral-weight redistribution inside the gap.

\begin{figure}[top]
\rotatebox[origin=c]{0}{\includegraphics[width=.95\columnwidth]{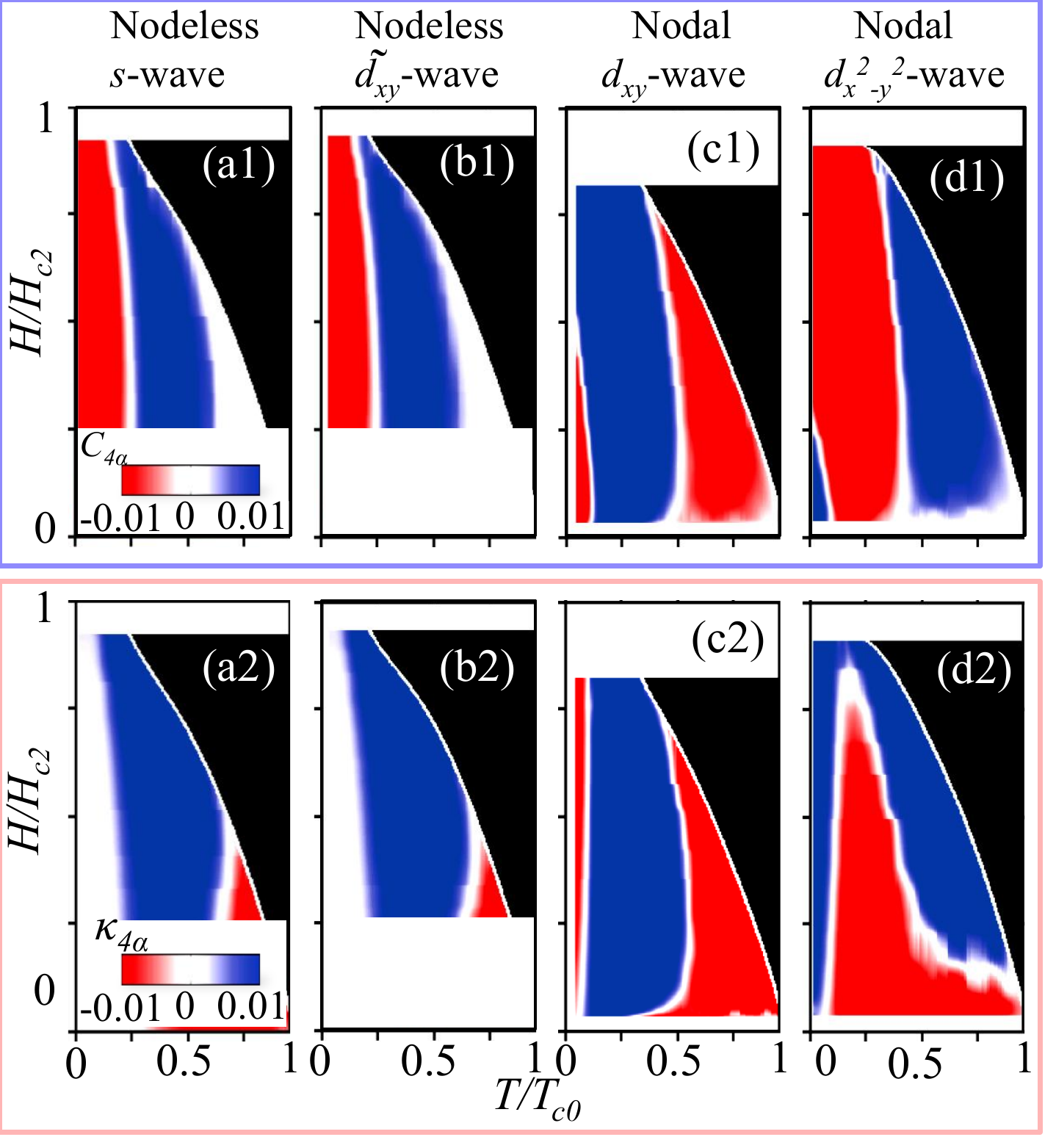}}
\caption{(Color online) Contour maps of fourfold amplitude oscillations of normalized specific heat, $C_{4\alpha}$ (top row), and normalized thermal conductivity $\kappa_{4\alpha}$ (bottom row) in the $H$-$T$ phase diagram. Each column denotes a different gap symmetry. All plots use the same color map (red to blue); a uniform color map is used for values below -0.01 and above 0.01. Note the fourfold amplitude is given with respect to $H//(100)$, i.e., a negative value corresponds to a minimum at $\alpha=0^o$.  Here $T_{c}({\bm H})$ is defined by the vanishing of both gaps for each  symmetry, which determines the line of the upper critical field $H_{c2}$.
} \label{fig4}
\end{figure}

In Fig.~3, we show the contour map of the amplitude of the fourfold oscillations  extracted from the normalized $\gamma=C/T$ (top row) and $\kappa/T$ (bottom row) for two nodeless and two nodal gaps. Earlier calculation using quasi-cylindrical FSs showed that the specific heat oscillation simply changes sign between the $d_{xy}$ and $d_{x^2-y^2}$ symmetries, while the overall phase diagram remains very much the same between them\cite{AntonI}. After the inclusion of realistic  and material-specific FSs in this work, we find substantial quantitative differences in the location of the sign-reversal lines between these nodal gaps in Figs.~3(c1) and 3(d1), due to the interplay of the SC order parameter with the FS anisotropies. Hitherto unknown is the intriguing result of both low-$T$ (strong) and high-$T$ (weak) sign reversals in the fourfold oscillations of $C(\alpha)$ and $\kappa(\alpha)$ for isotropic gaps, at moderate and high magnetic fields, see Figs.~3(a) and 3(b). We verified for $s$-wave pairing that the high-$T$ sign reversal is robust and remains at nearly the same location for a single-band superconductor with identical FS, while the low-$T$ feature disappears. Similarly, the low-$T$ sign change does not exist for two-band models with similar (in-phase) angular variations of Fermi velocities.

The striking feature of the phase diagrams for the heat capacity for nodal and nodeless cases in the top panels of Fig.~3 is that they all look qualitatively similar in the sense that they all exhibit sign reversals around the same $T$ and $H$. The same is true for the phase diagram of the $\kappa_{4\alpha}$ term in the thermal conductivity. Furthermore, the sign of the fourfold oscillations can sometimes be different for $C_{4\alpha}$ and $\kappa_{4\alpha}$ in the same region of phase diagram, as in the high-$H$ and low-$T$ region and vice versa for $d_{x^2-y^2}$-pairing in Fig.~3(d). This suggests that field-angle studies of each quantity alone are insufficient to distinguish between pairing symmetries. A simultaneous study of both $C(\alpha)$ and $\kappa(\alpha)$, including the comparison of the complex angle-dependent profiles and $T$ behavior, is necessary to image the gap structure. Our field-angle-dependent results of the nodal $d_{x^2-y^2}$-wave gap in the $H$-$T$ phase diagram are in qualitative agreement with recent specific heat data of CeIrIn$_5$ \cite{cerhin5}.

{\it Conclusions.$-$} The main conclusion of our work is that a mere observation of oscillations and sign reversals in $C(\alpha)$ or $\kappa(\alpha)$, combined with the $H_{c2}$ anisotropy, is insufficient to identify the presence of nodes or minima in the gap, and their interpretations require detailed knowledge of the underlying FS anisotropy. For multiband systems, the situation is further complicated by the interplay between multiband FS anisotropies and multiple SC gaps in that substantial fourfold oscillations and sign reversals can occur even for purely isotropic gap. These results are robust in the region of $0.5 H_{c2} < H < H_{c2}$ for isotropic gaps, and its region of validity extends to lower field with increasing gap anisotropy. Our results suggest that not only realistic theoretical calculations including field-induced impurity effect on mutiband systems are necessary,\cite{Mishra} the field-angle measurements should also be compared with $T$-dependence of the penetration depth, specific heat, and residual electronic term of $\kappa/T$ measurements for the detection of pairing symmetry\cite{Arberg,Hirschfeld,Graf}. 
In fact, in other probes such as quasiparticle interference (QPI) pattern seen in scanning tunneling microscopy/ spectroscopy (STM/S),\cite{Hanaguri} the consideration of the field induced impurity effect and the inclusion of realistic FS anisotropy should play an equally important role in the interpretation of data\cite{Wang}.

\begin{acknowledgments}
We thank R.\ Movshovich and A. V. Balatsky for discussions and encouraging this study. This work is funded in part by the US DOE under Grants No. DE-AC52-06NA25396 (TD and MJG) and DE-FG02-08ER46492 (IV), by NSF Grant No. DMR 0954342 (ABV), and by the Office of Science (BES) with a NERSC computing allocation under contract No.\ DE-AC02-05CH11231.
\end{acknowledgments}

\section{Supplementary Material}

A detailed account of the extended Brandt-Pesch-Tewordt (BPT) quasiclassical approximation in the vortex phase has been given in a series of papers \cite{AHoughton1998,Anton,AntonI,AntonII,Anton2010}. In this supplementary material we summarize the salient ingredients of this approximation and the numerical details of our quantitative studies.

\begin{figure*}[top]
\rotatebox[origin=c]{0}{\includegraphics[width=2.\columnwidth]{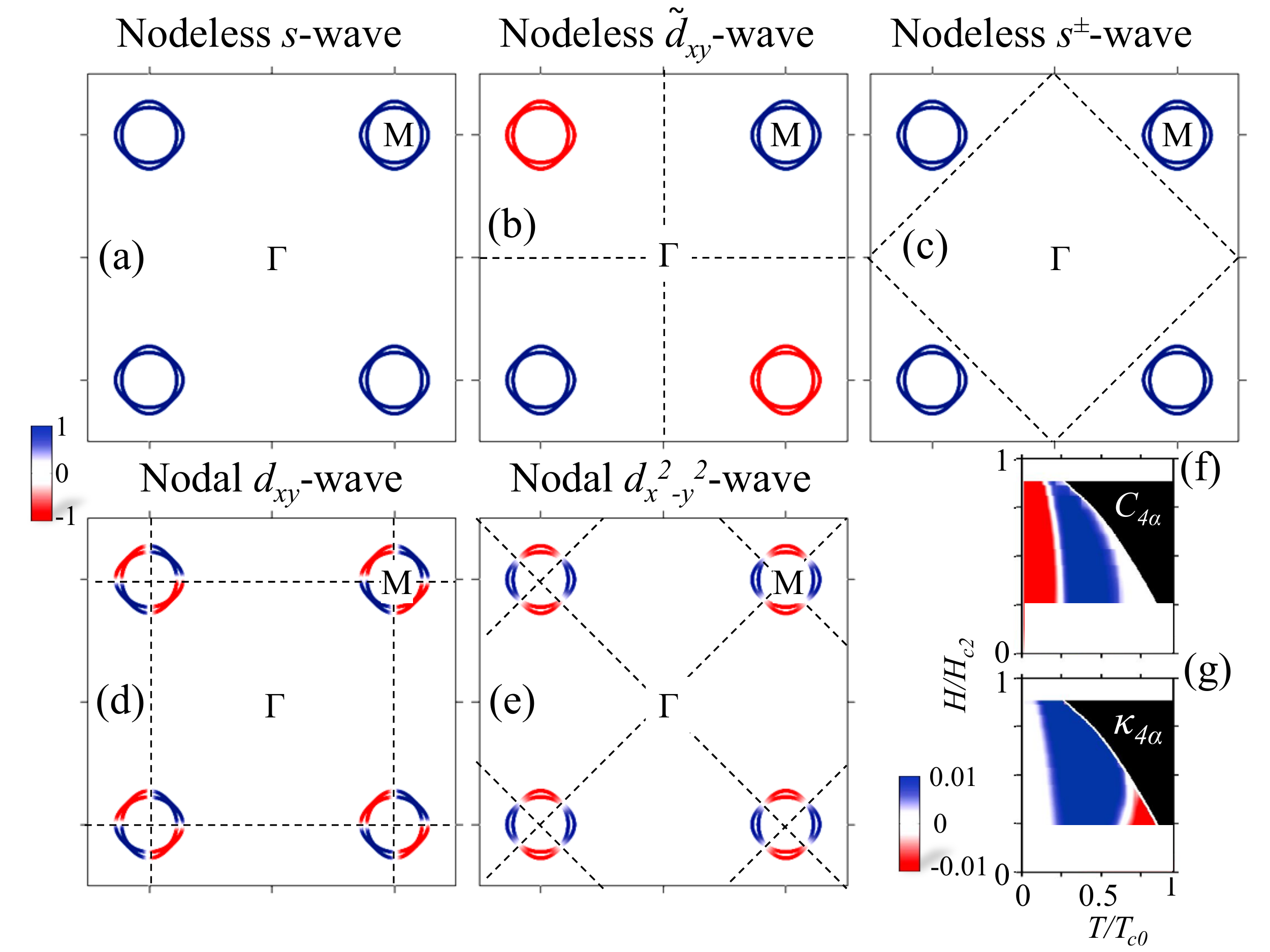}}
\caption{FIG.~S1. (Color online) Order parameters and phase diagrams for the model pairing states in $A_y$Fe$_{2-x}$Se$_2$. Panels (a)-(e): FS maps of the order parameter values for various pairing symmetries as indicated in each Panel. As shown in the color scale on the left, red means negative order parameter (OP), white denotes nodal regions,  and blue indicates positive OP. The dotted lines depict the nodal line. Panels (f)-(g): Phase diagram of the fourfold oscillations for $s^{\pm}$-pairing extracted from heat-capacity, $C_{4\alpha}$, and thermal conductivity, $\kappa_{4\alpha}$, respectively.} \label{fig4}
\end{figure*}

{\bf Fermi surface parameterization.$-$}
In order to calculate thermal properties in rotating fields, we use as input Fermi surfaces (FSs) and dispersions of $A_y$Fe$_{2-x}$Se$_2$ derived from first-principles electronic structure calculations to obtain an effective tight-binding model in the 2-Fe unit-cell notation given in Refs.~\cite{Das_FeSe,Dasmodulated}.  In accord with experimental observations, and essential for self-consistently calculated Green's functions for in-plane field rotation, a weak $k_z$ dispersion is added with hopping parameter $t_z\sim0.3t$,
where $t$ is the in-plane nearest neighbor hopping. Finally, a FS parameterization is obtained from the effect tight-binding model.
For each FS integration, we used 800 points along the Fermi line in the $(k_x,k_y)$ plane and 9 $k_z$ slices in the Brillouin zone. We confirmed numerical convergence by checking that the results do not change when we increase the number of $k$ points; this test was done for several points in the $T$-$H$ plane.

{\bf Multiband gap symmetries.$-$}
The superconducting (SC) pairing functions of all order parameters (OPs) considered for $A_y$Fe$_{2-x}$Se$_2$ are given in the main text. Fig.~S1 shows the two-dimensional view of of the corresponding OP on each Fermi surface sheet. The conventional $s$-wave gap is isotropic at all momenta as shown in  Fig.~S1(a). The extended $\tilde{d}_{xy}$ order parameter is nearly isotropic on each of the FSs, but changes sign between FS sheets located at different M and equivalent points in the Brillouin zone (BZ), see  Fig.~S1(b). This is one of the candidate order parameters for this system, which gives rise to a spin resonance \cite{Das_FeSe,Dasmodulated}. The $s^{\pm}$ gap has a nodal line at the mid-point between the $\Gamma$ to M direction, but that nodal line does not touch either FS. Importantly, due to the absence of a hole pocket at the $\Gamma$ point, there is no sign change in the SC gap in  $A_y$Fe$_{2-x}$Se$_2$ for this pairing. The nodal line passes through the zone boundary and zone diagonal for $d_{xy}$ and $d_{x^2-y^2}$-pairings as presented by dashed lines in Figs.~S1(d) and S1(e).

{\bf Computational details.$-$}
For each pairing symmetry, the coupled order parameters are computed self-consistently at each given magnetic field (${\bm H}$ applied at angle $\alpha$ to the (100) direction, and temperature ($T$). To plot the phase diagram in Fig.~3 in the main text, we took a mesh of 35 field points between zero and $H_{c2}$, 100 temperature points from zero to $T_{c0}$, and used 31 $\alpha$ points from zero to $90^o$ to extract the anisotropic terms in the heat capacity and the thermal conductivity. The phase diagram for $s^{\pm}$ pairing is very similar to that of the nodeless $s$ and $\tilde{d}_{xy}$ gaps, see  Fig.~S1(f)-S1(g), and thus not included in the main text. For all calculations,  we considered  purely intraband impurity scattering in the clean limit, $1/\tau_{imp}=0.01 \times 2\pi T_{c0}$, where $T_{c0}$ is the bare transition temperature and the scattering phase shift is chosen to be $\delta=\pi/2$  (unitarity limit).

{\bf Methods and formulas.$-$}
If the Fermi velocity in band $n$ is denoted by $\bm{v}_n({\bm k_f})$, the corresponding normal-state DOS at the Fermi level is $N_{fn}({\bm k_f})\sim1/|\bm{v}_n({\bm k_f})|$. The wavevector ${\bm k_f}$ lies on the respective FS.
When a magnetic field ${\bm H}$ is applied along an angle $\alpha$ with respect to the $x$-axis, the relevant parameter in the SC state is the  component of the Fermi velocity normal to the applied field, which, in energy units, becomes $\bm{\bar{v}}_{fn}(\phi,{\bm H})\equiv {\bm v}_n^\perp({\bm k_f})/2\Lambda$,
where $\Lambda=(\hbar c/2|e|H)^{1/2}$ is the magnetic length of order of the intervortex distance,  $\phi$ is the FS angle with respect to the $k_x$ axis,
and $\bm{v}_n^\perp({\bm k_f})$ is the rescaled component of the Fermi velocity perpendicular to ${\bm H}$.  The unit-cell averaged Green's function at the fermionic Matsubara frequency, $\omega_\nu$,
follows the notation of Eqs.~(46)-(48) in Ref.~\onlinecite{AntonI}:
\begin{equation}
g_n(i\widetilde\omega_{\nu}, {\bm k_f}; {\bm H}) =
\frac{-i\pi}{
 \sqrt{ 1-\left(\frac{\widetilde\Delta_n({\bm k_f}; {\bm H})}{i\widetilde\omega_\nu-\Sigma_n(i\widetilde\omega_\nu,{\bm k_f}; {\bm H})}\right)^2}}.
\end{equation}
Here $\widetilde\Delta_n$ and $\widetilde\omega_{\nu}$ are the order parameter and Matsubara frequency renormalized by the impurity self-energy, which is evaluated in the $T$-matrix approximation.
The self-energy  $\Sigma_n$ is 
given by $(i\widetilde\omega_\nu-\Sigma_n)^{-2}=i\sqrt{\pi} W_n^{\prime}(i\widetilde\omega/\bar{v}_{fn})$.
$W^{\prime}(z)$ is the first derivative of the complex-valued function $W(z)=\exp{(-z^2)}{\rm erfc}(-iz)$. In contrast to the Doppler shift approximation, both the real and the imaginary parts of $\Sigma_n$ contribute to the SC DOS, and their interplay as a function of energy, $H$ and $T$, 
determine the sign reversal in the fourfold oscillation of the SC DOS.
These effects have been extensively studied earlier using a single quasi-cylindrical FS and nodal gap, and a minimal 2D model for two-band systems, see for example Refs.~\onlinecite{Anton,AntonI,AntonII,Anton2010}.

The order parameters are calculated self-consistently from the coupled gap equations of the two-band model
\begin{eqnarray}
&&\Delta_n({\bm k_f};{\bm H}) =  \nonumber\\
&&
T \sum_{\omega_\nu} \sum_{n'}
\Big\langle V_{n n'}({\bm k_f}, {\bm k_f}')
N_{fn'}({\bm k_f}')  f_{n'}(i\omega_\nu, {\bm k_f}'; {\bm H}) \Big\rangle_{FS} .
\end{eqnarray}
Here $f_{n'}$ is the anomalous Gorkov function (off-diagonal Green's function).
We used a factorized pairing potential at the Fermi surface as $V_{n n'}({\bm k_f}, {\bm k_f}') = V_{n n'} {\cal Y}_n(\phi) {\cal Y}_{n'}(\phi')$, with ${\cal Y}_n(\phi)$ the azimuthal angle dependence of the SC gap, see Fig.~1 in main text. For simplicity, we consider purely interband pairing $V_{12}=V_{21}=-V$ and eliminate $V$ in favor of the bare transition temperature $T_{c0}$ using weak-coupling theory \cite{Anton2010}.

The specific heat, $C=C_1+C_2$, and thermal conductivity, $\kappa=\kappa_1+\kappa_2$, are calculated
from the approximate expressions
\begin{eqnarray}
&&C_n(\alpha) \! \approx \!\!
\int_{-\infty}^\infty \!
d\omega \frac{\omega^2 \langle N_n(\omega, {\bm k_f}; {\bm H}) \rangle_{FS}}{4 T^2 {\rm cosh}(\omega/2T)^2} ,
\label{eq:C}
\\
&&\kappa_{n}^{xx}(\alpha) \! \approx \!\!
\int_{-\infty}^\infty \!\!
d\omega \frac{ \omega^2 \langle v_{n}^x({\bm k_f})^2 N_n(\omega, {\bm k_f}; {\bm H})\tau_{n}(\omega,{\bm k_f}; {\bm  H}) \rangle_{FS} }{2 T^2 {\rm cosh}(\omega/2T)^2},
\nonumber\\
\end{eqnarray}
We note that close to the transition the full expression for the heat capacity computed from the entropy includes the temperature derivative of the gap functions $\Delta_n$. However,  inclusion of these terms only minimally affects quantitative aspects of the results away from the transition.
The field-induced SC DOS in each band, $N_n/N_{nf}=-{\rm Im}\ g_n/\pi$, is calculated using analytical continuation $i\omega_\nu\rightarrow\omega+i\delta$ to obtain the retarded Green's functions, and
the {\it transport} lifetime due to both impurity and vortex scattering \cite{Vekhter99,AntonII,Anton2010}:
\begin{widetext}
\begin{eqnarray}
  \frac{1}{2\tau_{{n}} (\omega,{\bm k_f}; {\bm  H})} =
    - {\rm Im}\, \Sigma_n(\omega,{\bm k_f};{\bm H}) +
    \sqrt{\pi}{1 \over |{\bar{\bm v}}_{fn}({\bm k_f}; {\bm H})|}
    \frac{{\rm Im}\, [g_n(\omega,{\bm k_f};{\bm H}) \, W(\tilde{\omega}/|{\bar{\bm v}}_{fn}({\bm k_f}; {\bm H})|)]}
    {{\rm Im} \, g_n(\omega,{\bm k_f};{\bm H})} |\Delta_n({\bm k_f}; {\bm H})|^2
    \,.
    \label{eq:tauH}
\end{eqnarray}
\end{widetext}
Since the function $x^2/{\rm cosh}(x/2)^2$ has a peak at $x\sim 2.5T$, the heat capacity at low temperatures predominantly probes the anisotropy in the total SC DOS, $N(\omega=2.5T,{\bm k_f}; {\bm H})$. Using the expansion of  the error function, we obtain two limiting values for $W^{\prime}(z)$: $W^{\prime}(0)=2i/\sqrt{\pi}$ and $W^{\prime}(z\gg 1)\approx-i/\sqrt{\pi}z^2$. Thus the SC DOS for each band $n$ becomes
\begin{widetext}
\begin{eqnarray}
N_n(\omega; {\bm H}) = \langle N_n(\omega,{\bm k_f};{\bm H})\rangle_{FS}
\approx
\begin{cases}
\left\langle N_{fn}({\bm k_f})\left[1+\left(\frac{\widetilde\Delta_n({\bm k_f}; {\bm H})}{|\bar{\bm v}_{fn}({\bm k_f}; {\bm H})|}\right)^2\right]^{-1/2}\right\rangle_{FS} ,
&\omega\ll \bar{\bm v}_{fn} , \cr\nonumber
\left\langle N_{fn}({\bm k_f})\left[1- \left(\frac{\widetilde\Delta_n({\bm k_f}; {\bm H})}{\widetilde\omega}\right)^2\right]^{-1/2}\right\rangle_{FS} ,
&\omega\gg \bar{ \bm v}_{fn}  ,\cr
\end{cases}\\
\label{Eq:N}
\end{eqnarray}
\end{widetext}
where $\widetilde\Delta_n$ and $\widetilde\omega$ are the impurity renormalized order parameter and quasiparticle energy, respectively.
The first line above only makes physical sense when the BPT approximation is valid at low energies, i.e., for nodal gaps. In that case at low $T$ (low energy) and at low fields, where $\Delta_n({\bm k_f}; {\bm H})$ only weakly depends on the direction of the field, the SC DOS depends predominantly on the orientation of
$|{\bar{ \bm v }}_{fn}({\bm k_f};{\bm H})|$ relative to the minima of $\Delta_n({\bm k_f}; {\bm H})$. At $\omega=0$ the inversion of the SC DOS as a function of the field for nodal gaps can be obtained in analogy to Refs.~\cite{UdagawaI,AntonI}.

At higher energies, the second line of Eq.~\eqref{Eq:N} gives the BCS result for the SC DOS and therefore field-angle variation enters via the anisotropy of the upper critical field that influences $\Delta_n({\bm k_f}; {\bm H})$ in the vicinity of the transition. This result is valid for both nodal and nodeless gaps, including the fully isotropic situation. Crucially, for anisotropic Fermi surfaces the $H_{c2}$ anisotropy in the order parameter is weighted by the normal-state DOS, $N_{fn}({\bm k_f})$, leading to a complex behavior including the switching of the minima and maxima found in our Letter. In this regime, however, the energy width of the Fermi weighting factor in the integral exceeds the gap amplitude and a full numerical evaluation is required. The results of such a self-consistent analysis are presented in the main text.  All our results are consistent with the general observations based on this expansion.


\begin{thebibliography}{10}
\bibitem{impurity} Here, we consider the clean limit with purely intraband impurity scattering rate $1/\tau_{imp}=0.01 \times 2\pi T_{c0}$ and strong scattering with phase shift $\delta=\pi/2$  (unitarity limit). The pure transition temperature is $T_{c0}$.
\end{thebibliography}
\end{document}